\documentclass[journal]{IEEEtran}
\usepackage{cite}
\usepackage{graphicx}
\usepackage{amsmath,amssymb}
\usepackage{tabularx}
\usepackage{color}
\usepackage{comment}
\usepackage{url}
\usepackage{bm}
\usepackage{mdframed,xcolor}
\def \am [#1]{\textcolor{red}{AM: #1}}
\def \rev [#1]{\textcolor{black}{#1}}
\def \revv [#1]{\textcolor{blue}{#1}}

\title{Sound Event Detection: A Tutorial}
\author{
Annamaria Mesaros$^1$, Toni Heittola$^1$, Tuomas Virtanen$^1$, Mark D. Plumbley$^2$
\thanks{This work was supported in part by the European Research Council under the European Unions H2020 Framework Programme through ERC Grant Agreement 637422 EVERYSOUND,
Grants EP/N014111/1 "Making Sense of Sounds" and EP/T019751/1 "AI for Sound" from the UK Engineering and Physical Sciences Research Council (EPSRC), and 
Academy of Finland grant 332063 "Teaching machines to listen".}\\
$^1$ Tampere University, Finland \\
email: \{annamaria.mesaros, toni.heittola, tuomas.virtanen\}@ tuni.fi \\
$^2$ University of Surrey, UK \\
email: m.plumbley@surrey.ac.uk
}

\begin{document}

\maketitle

\section{Sound events in our everyday environment}

Imagine standing on a street corner in the city. With your eyes closed, you can hear and recognize a succession of sounds: cars passing by, people speaking, their footsteps when they walk by, and the continuously falling rain. Recognition of all these sounds and interpretation of the perceived scene as a city street soundscape comes naturally to humans. It is, however, the result of years of ``training":  encountering and learning associations between the vast variety of sounds in everyday life, the sources producing these sounds, and the names given to them.

Our everyday environment consists of many sound sources that create a complex mixture audio signal. Human auditory perception is highly specialized in segregating the sound sources and directing attention to the sound source of interest. This phenomenon is called \textit{cocktail party effect}, as an analogy to being able to focus on a single conversation in a noisy room.
Perception groups the spectro-temporal information in acoustic signals into \emph{auditory objects} such that sounds or groups of sounds are perceived as a coherent whole \cite{Gaver1993}. This determines for example a complex sequence of sounds to be perceived as a single sound event instance, be it ``bird singing" or ``footsteps". 

\begin{figure*}
    \centering
    \includegraphics[width=\textwidth]{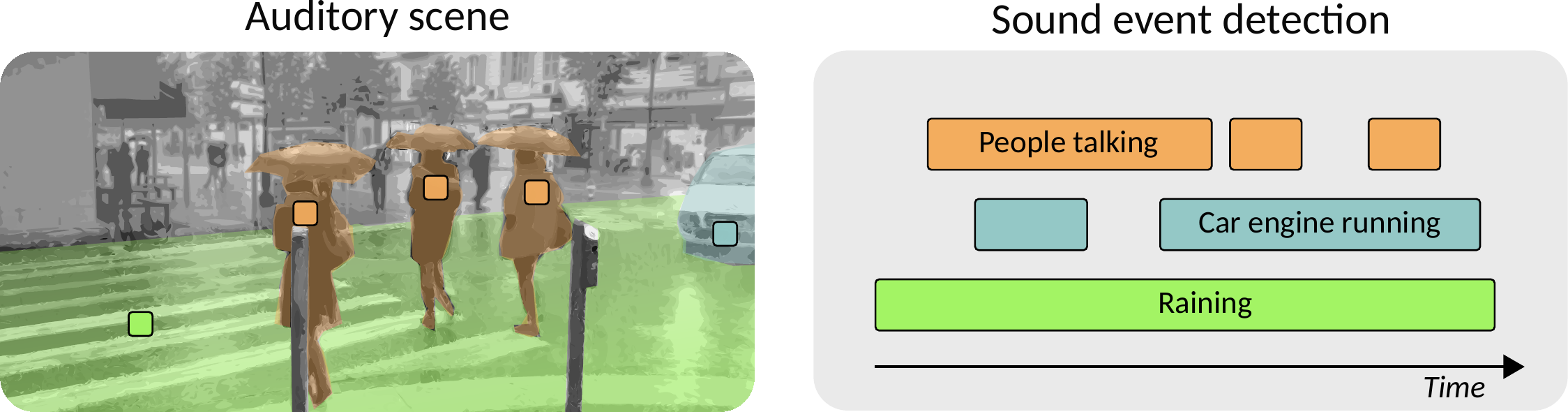}
    \caption{Sound event detection: find the temporal activity of sound events in the auditory scene, and output the class and temporal information on all such instances.}
    \label{fig:sed-definition}
\end{figure*}

The goal of automatic sound event detection (SED) methods is to recognize \textit{what} is happening in an audio signal and \textit{when} it is happening. In practice, the goal is to recognize at what temporal instances different sounds are active within an audio signal. This task is illustrated in Fig.~\ref{fig:sed-definition}.

The set of target sounds for a detection task are specific to each application, but in the case of a general-purpose sound event detection system the target sounds are environmental sounds such as bird singing, car passing by, and footsteps. In the literature, these are sometimes referred to as \textit{non-speech and non-music sounds}\cite{Gygi_2007}, to differentiate the field of environmental sound analysis from more established speech or music analysis tasks. The sound event detection task also has a different purpose than the typical speech or music analysis tasks, because perception of speech, music, and environmental sounds is also different: while musical listening focuses on the aesthetic qualities of the sound, and speech perception focuses on the linguistic or paralinguistic information, \textit{everyday listening} is directed towards identification of the sound sources \cite{Gaver1993}.

\section{Challenges in sound event detection}

Creating automatic systems for sound event detection is hindered by multiple challenges, some related to the nature of the sounds to be detected and how they appear in natural environments, others related to data collection and annotation procedures. In turn, these determine the challenges that must be overcome by the machine learning techniques in the learning process. 

Sound events have very diverse acoustic characteristics. Some sounds are very short, transient-like, others may have longer duration and be harmonic, while events such as ``footsteps" or ``bird singing" may arise from temporal integration of a sequence of elementary components. In typical applications of sound event detection, the target sound events are far away from the microphone, so the events are significantly affected by the acoustic transfer function. Also because of the distance to the microphone, the sound pressure level of the target sound event when received by the microphone can be often lower than that of other sounds occurring in the environment, adding to the difficulty of detection.

Natural environments are polyphonic, meaning that multiple sounds may be active at the same time. Polyphony exists also in music, although often in a constrained way where overlapping sound events exhibit harmonic relations with each other, so that fundamental frequencies form small integer ratios \cite{benetos2019}.
For sound events, there are no predefined rules on how sounds can co-occur, and the way to model co-occurrence is through statistics from the data itself, such as direct counts, and degree of overlap  of different classes in the data.

The number of possible sound event classes is unlimited, since any object or being may produce a sound as a  naturally-occurring event \cite{Gygi_2007}. This is entirely different from other audio classification domains such as speech recognition, where a fixed set of classes (phonemes) is possible in a language, and language models can provide support to acoustic models in the recognition task. 

The case is further complicated by the lack of an established ontology for universally describing sound classes. Categories of sounds are often defined according to some common feature like the source or the production mechanism \cite{guastavino2018everyday}, but this allows a high degree of ambiguity in the category definition. 

In practice, each application of sound event detection is targeting slightly different sound classes, and is used in different environments. Given all this diversity, there are no universally applicable sound event detection datasets or models, but each application requires data collection and system development to meet its specific needs. 
This is a considerably different approach than for speech recognition, where we expect to develop a system able to cope with all combinations of sounds.

\section{The general machine learning approach for sound event detection}
\label{sec:general-ml}

The dominant approach to tackle the sound event detection task is based on \textit{supervised learning} \cite{Mesaros2018_TASLP}, where a training set of audio recordings and their reference annotations of class activities are used to learn an \emph{acoustic model}. The annotations contain information about the temporal activity of each target sound class in a binary manner, describing whether a class is active or not, for each time unit. This representation is sometimes referred to as an \textit{event roll}, after the piano roll representation \cite[Chapter~1]{muller2015fundamentals} in music information retrieval. 

\begin{figure*}
    \centering
    \includegraphics[width=1.0\textwidth]{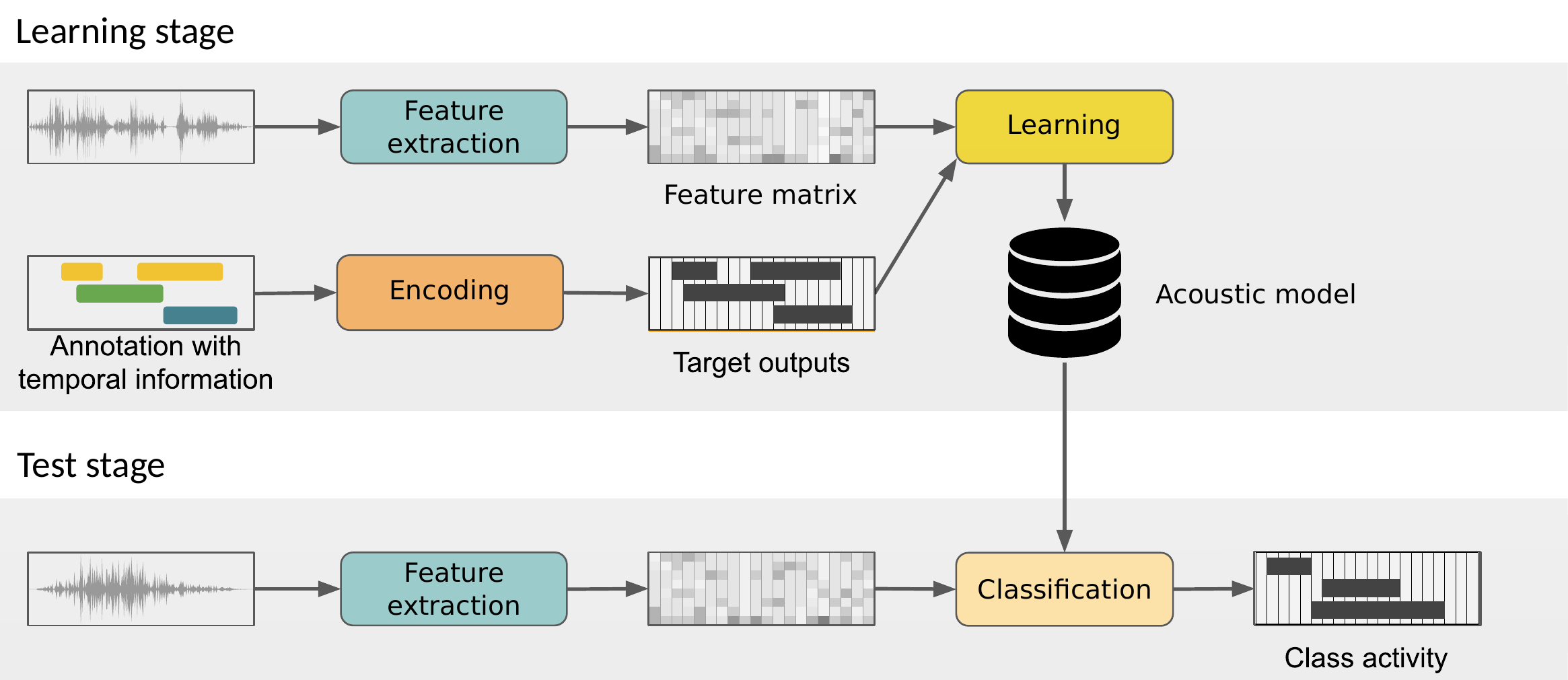}
    \caption{Overview of a sound event detection system, where activities of classes in short consecutive segments of audio are estimated by multi-label classification.}
    \label{fig:general_system}
\end{figure*}

A general classification system illustrating supervised learning for sound event detection is presented in Fig.~\ref{fig:general_system}. 
To detect multiple sound event classes, the system must provide an output for each considered class, in order to indicate if a sound of that class is active. In the learning stage, the system learns the correspondence between the features extracted from the audio signal and an annotation that represents the activity of each class. The annotations are represented as a binary matrix in which each element represents a class being active (1) or not active (0) within short temporal segments. At test stage, the system receives the features extracted from a test audio recording, and provides an output represented according to the same convention: a matrix indicating binary activity for each sound class in consecutive time segments. In general, a single sound event detection system is used to predict activities of multiple sound classes which can be active simultaneously, leading to \textit{multi-class multi-label classification} in each segment. 

\begin{figure*}
    \centering
    \includegraphics[width=0.9\textwidth]{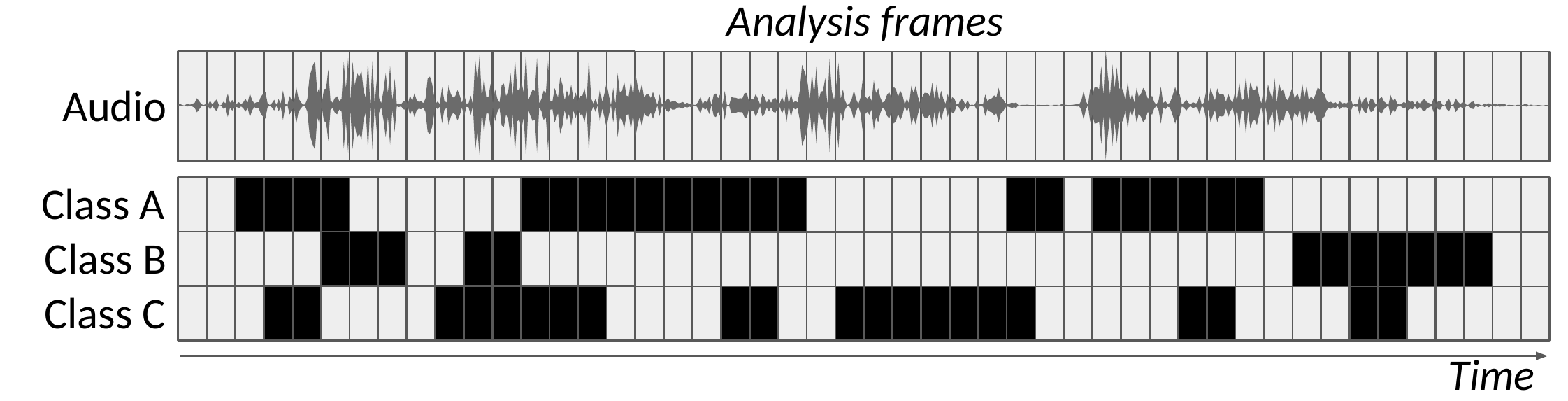}
    \caption{Sound event detection as multi-label classification of short consecutive audio segments.}
    \label{fig:sed}
\end{figure*}

Sound event detection aims to estimate the \emph{temporal} activity of sounds in an audio recording---i.e., a label does not necessarily apply to the entire recording, but the temporal activity pattern for a label is estimated within it to find individual sound instances. The multi-class multi-label classification in consecutive temporal segments of the same audio example results in an output such as the one illustrated in Fig.~\ref{fig:sed}. The lengths of these segments are determined by the target output resolution, and can be as short as the analysis audio frames (typically 20-100 ms) or longer. The analysis frames define the length of audio in which the features are calculated, while the system can perform prediction at frame-level or at different resolution. When the system provides a single prediction per class for the entire duration of the audio file, and does not output separate activity patterns for each sound, 
the task is no longer referred to as detection, but rather as tagging.

Context information from consecutive analysis frames brings more information to learning contiguous segments corresponding to sound event instances. Temporal context can be modeled using different techniques depending on the classification approach, and will be discussed later. Also the system output in consecutive segments may be postprocessed in order to transform the information into sound event instances. For example one or a few consecutive segments in which a sound was detected as active may still have a combined length that is too short to be a plausible event instance, so they would be discarded. Similarly, short gaps in event activity may be ``filled in" under the assumption that multiple fragments form a single event instance. The expected minimum/maximum duration of sound instances can be based on statistics obtained from the training data or on general assumptions about the target sounds (e.g. 500~ms is too short for ``car passing by").

Early approaches for sound event detection borrowed techniques from speech recognition and music information retrieval, and were therefore based on traditional pattern classification techniques like Gaussian mixture models (GMMs) and hidden Markov models (HMMs). 
However, such models are much more useful in speech and music modeling due to specific techniques that allow modeling of elementary units in speech or music, like state-tying of phonemes or left-to-right topologies for modeling temporal evolution of phonemes and musical notes. Sound events in general do not consist of similar elementary units as speech, making GMMs and HMMs less relevant for SED.
In addition, these methods are not designed to detect multiple classes at the same time. In order to solve the multi-label classification they required specific extensions or setups like for example binary classification for each sound event class \cite{Mesaros2016_EUSIPCO}, or multiple passes of Viterbi decoding \cite{heittola2013_EURASIP} or preprocessing involving sound source separation \cite{Heittola2011_CHIME}. In contrast, modern pattern classification tools, especially deep neural networks (DNNs), can perform multi-label classification more easily: multiple output neurons being active at the same time indicate activity of multiple sound classes at the same time. 
This gives DNNs a tremendous advantage in solving the multi-label classification problem, and has made them predominant in the field. With DNNs, sound event detection has seen major leaps in state-of-the art performance and the complexity of tackled problems.

\section{Data}

In order for SED methods developed using supervised learning to operate well, they need to be trained on audio data from conditions that are similar to the intended usage scenario, with annotations of the target sound events. Application that use sound event detection often have different sets of target sound event classes and may be used in slightly different environments. Therefore, there is no universally applicable acoustic model or sound event detection dataset, and instead many datasets are collected for the problem at hand.

Collecting audio data in everyday environments requires planning of the situations to be recorded. A sufficiently large number of examples of each target sound class need to be recorded or acquired from existing sources, in order to cover \textit{acoustic variability} for the class. This variability comes from different factors. For example in the case of a ``dog barking" class, this includes different instances of the source (different dogs), differences in the source state (for example mood), and environmental factors such as the space where the source is and what other sound sources are around and active, the location of the source, and the position of the microphone used to capture the sound. There is no rule for how many examples and situations are needed: depending on the complexity of the problem, sufficient data is needed for robust training of acoustic models.

Many environmental factors can be simulated to some degree, for example by adding interfering sources to an audio signal or by convolving the signal with various room impulse responses. Nevertheless, many environmental sounds (for example a car passing by) are difficult to capture in isolation, making it difficult to control these factors in an audio recording. Therefore, often the only way to acquire training data is to capture a diverse set of training examples of the target sources in realistic environments. In practice, it is impossible to cover all variability of a source class, but a sufficient diversity of the training data can be expected to lead to good  generalization of the trained models.

\begin{figure*}
    \centering
    \includegraphics[width=0.8\textwidth]{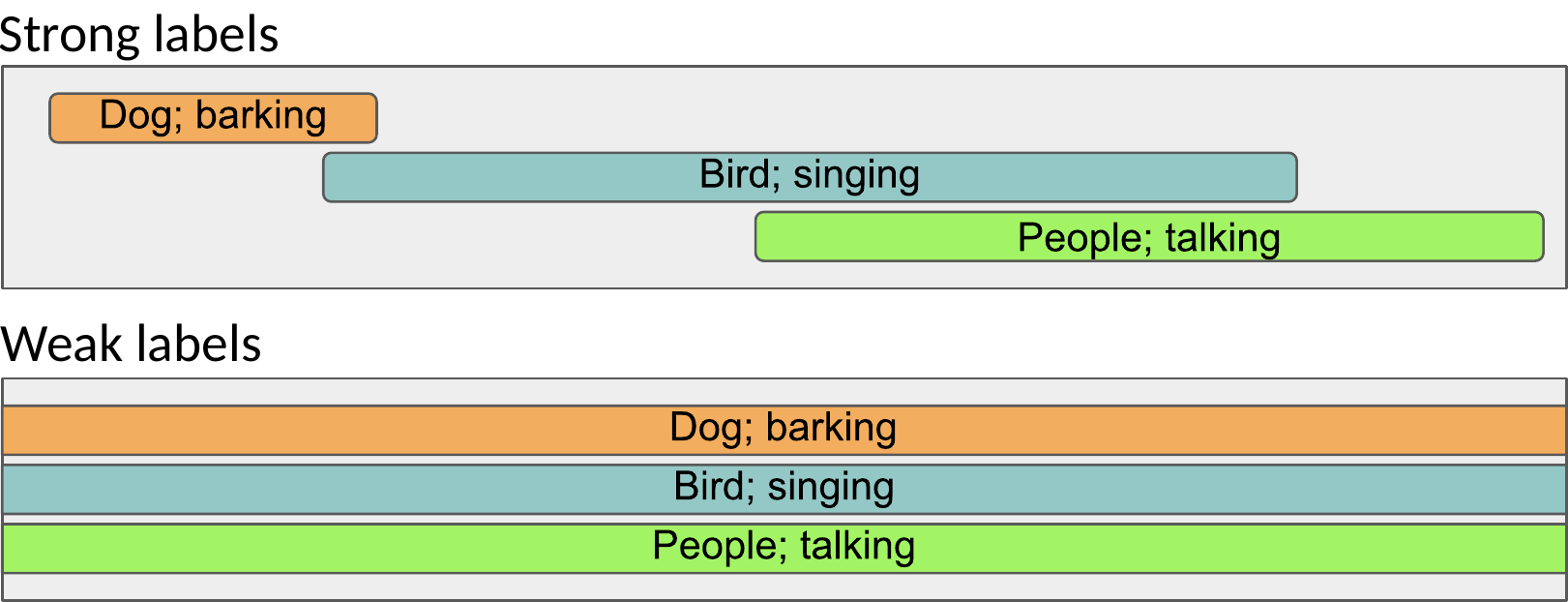}
    \caption{The main difference between strong and weak labels: a strong label indicates the exact temporal segments of the audio recording where the event is active, while a weak label indicates only that the sound event is active within the recording.}
    \label{fig:annotation_process}
\end{figure*}

For setting up a multi-class multi-label classification problem in short time segments, the annotation of the audio must contain information about the sound events in short time segments too. 
This type of annotation is a very specific requirement that is referred to as \textit{strong labels}, i.e., the annotation contains temporal information for each sound instance, namely its onset and offset times. On the other hand, a \textit{weak label} informs only about the presence of the sound in a longer audio recording, without explicitly indicating the temporal region where this sound is active. In other words, a weak label applies to the entire audio recording, while a strong label applies to specific segments of the audio, as illustrated in Fig.~\ref{fig:annotation_process}.
Ideally, the temporal precision of strong labels  used to train the system should be at least as good as the output resolution. However, in practice one may resort to lower resolution or weak labels in order to allow fast production of large quantities of training data.

The set of target sound classes is dictated by the application for which the system is developed. A naive approach to data collection is to try to annotate all audible sounds, by selecting the segments where they are active and giving them a detailed label \cite{mesaros2018datasets}. Such an approach may result in a set of very diverse labels, with an insufficient number of examples to train an acoustic model for some classes.  
For this reason, a good policy is to select the target classes in advance and set some rules for their format, for example requiring a noun+verb pair indicating the source and the action producing the sound \cite{guastavino2018everyday, Mesaros2016_EUSIPCO}. Manual annotation of onsets and offsets is also difficult in polyphonic mixtures, making the annotation process slow when strong labels are needed. On the other hand, because weak labels do not contain any temporal information, annotation with weak labels can be considerably faster. 

Labeling can be done either by expert annotators or using \textit{crowdsourcing}. Crowdsourcing by non-expert annotators is suitable for weak labeling, but is prone to errors due to factors such as lack of expertise or carelessness in selecting a label that matches the heard content. Crowdsourced annotations may be referred to as \textit{ noisy data} \cite{fonseca2019learning}, a term which includes cases when the label is incorrectly attributed or missing \cite{fonseca2020missinglabels}. Noisy data can be curated for improving its quality, or used directly as is. Learning from noisy or missing labels requires specific machine learning techniques or cost functions that are able to compensate for the unreliability of the labels \cite{fonseca2019learning, fonseca2020missinglabels}. 

The most reliable way of creating strong labels is to generate synthetic audio signals by mixing individual sound event instances with a background audio signal which may be recorded in a specific environment \cite{Mesaros2018_TASLP}, or simply noise \cite{Salamon2017}. This allows fast production of accurate annotations, since the start and end times of sound events are determined by the mixing algorithms \cite{Salamon2017}. Generating synthetic mixtures also allows the simulation of a wide variety of situations, including different signal to noise ratios, different backgrounds, and different properties of the acoustic space. On the other hand, current synthesis methods have no rules for selecting the sound event combinations, and the generated data may result in unnatural combinations, or fail to capture the complexity of real-life environments. Methods trained using only synthetic data will most likely not perform equally well on data from real environments. Using advanced enough synthesis procedures can lead to large amounts of diverse data, which may provide at least some degree of generalization.

Table \ref{tab:datasets} includes a few examples of publicly available datasets, including their statistics and annotation method. It is clear even from this small number of examples that the size of the dataset is closely determined by the difficulty of annotating it: manually annotated data is smaller, with the strongly-labeled TUT Sound Events 2016 dataset \cite{Mesaros2016_EUSIPCO} being much smaller than the weakly-labeled and crowdsourced FSDnoisy18k \cite{fonseca2019learning}. At the other extreme is AudioSet \cite{Gemmeke2017}, a large dataset with a high number of classes, having annotations obtained through automatic methods and partial human verification. The synthetic data example, URBAN-SED \cite{Salamon2017}, is based on a set of 10 target event classes and released as a benchmark dataset, but the generation procedure is publicly available and can be used for generating datasets with any other target sound event classes. 

\begin{table*}[]
    \centering
    \caption{Datasets used for sound event detection and classification }
    \begin{tabular}{l|lll|cr|l}
         Dataset & Audio source & Annotation & Labels & Classes & Data & Notes\\
         \hline
         TUT Sound Events 2016 & Recordings & manual & strong & 18 & 2h & Real-life recordings\\
         URBAN-SED & Synthetic & synthetic & strong & 10 & 30h & Synthetically generated material \\        
         AudioSet & YouTube & automatic & weak & 527 & 5000h & Automatically tagged videos,\\
         & & & & & & partially verified\\
         FSDnoisy18k & Freesound & crowdsourced & weak & 80 & 90h & Curated / verified annotations and \\
         & & & & & & noisy crowdsourced annotations \\
    \end{tabular}
    \label{tab:datasets}
\end{table*}

\section{Signal processing methods for sound event detection}

At a general level, the sound event detection task involves two main stages: feature representation and classification. The classification has a training phase in which the system learns the acoustic models, and a test phase in which the acoustic models are used to provide predictions on test data.
In addition, the system may involve some pre-processing, for example, to remove noises from the sound. At training stage, a process called  \emph{data augmentation} is often used to produce more training data and increase its diversity.

\subsection{Data augmentation}

Training data is sometimes difficult to obtain, so data augmentation is a way to artificially increase the amount and diversity of training data. When real-life recordings are available, the combinations of overlapping sounds cannot be controlled, therefore many reasonable combinations of sounds will be absent from the training data. This is a specific characteristic of the polyphonic sound event detection problem, and is different from analysis and recognition of audio from speech or music. For audio containing single sound sources like speech from a single speaker, data augmentation may be used for supplementing acoustic conditions such as noise or room characteristics. 
On the other hand, in polyphonic music the overlapping sounds often exhibit harmonic relations with each other, and data augmentation should retain this property in order to create plausible additional data. 
In sound event detection, manipulating and combining the available audio allows the creation of new combinations of overlapping events. In addition, multiple noise and impulse response conditions can be used to supplement the data through mixing and convolution with the original data.  All these techniques contribute to the robustness of the acoustic modeling process, by adding acoustic variability to the training data. 

\begin{figure*}
    \centering
    \includegraphics[width=\textwidth]{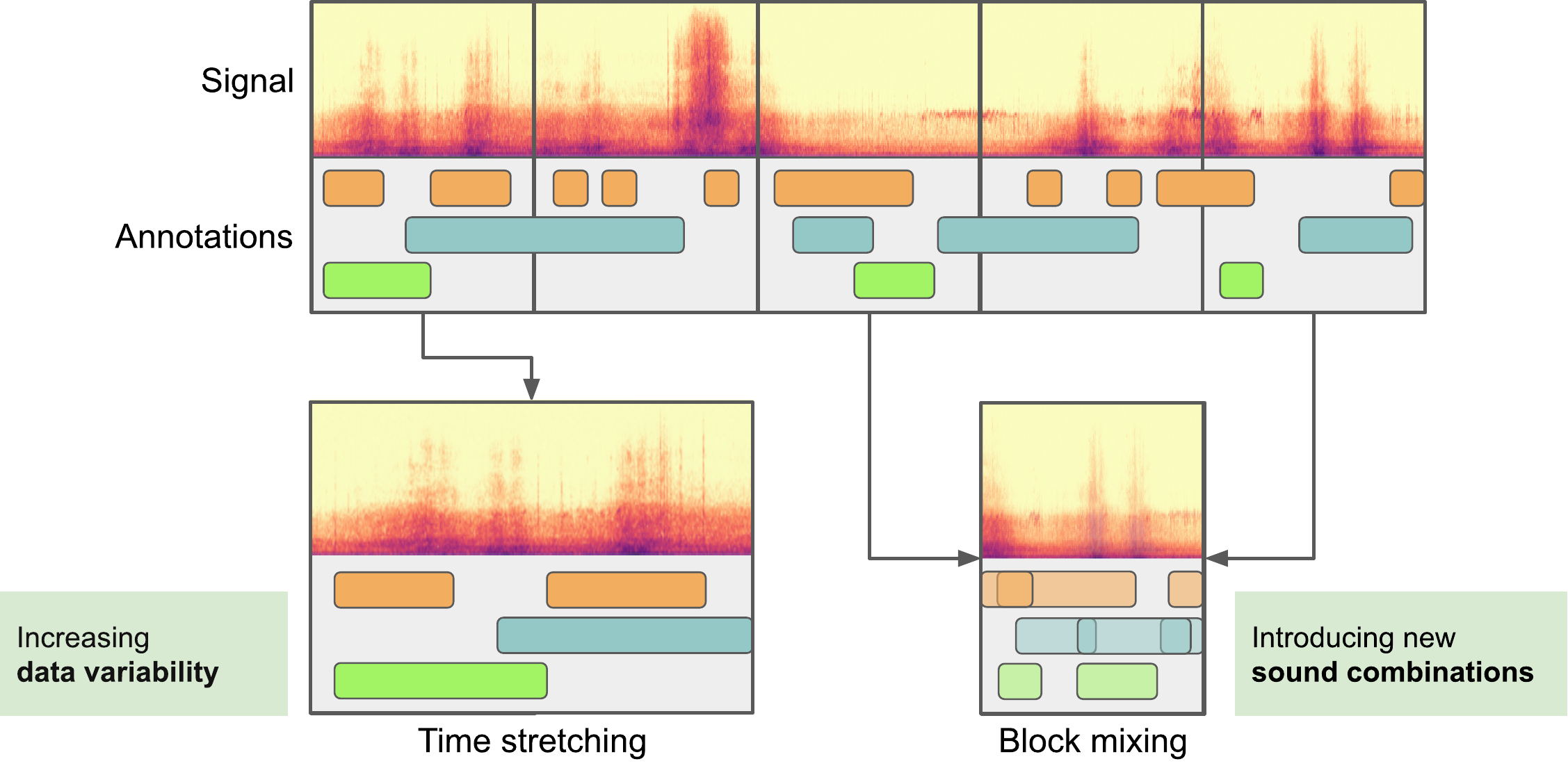}
    \caption{Data augmentation using time stretching and block mixing introduces acoustic variability and new sound combinations.}
    \label{fig:data-augmentation}
\end{figure*}

Methods for data augmentation used in SED range from basic signal manipulations such as time stretching, pitch shifting, and dynamic range compression \cite{Salamon2017}, to more complex ones such as convolution with various impulse responses (to simulate various microphones and acoustic environments) \cite{politis2020dcase}, sub-frame time shifting, block mixing \cite{parascandalo2016}, mixup \cite{Dorfer2018}, and simulating a set of noise conditions by adding background noise while varying the SNR \cite{Salamon2017}.  

When performing data augmentation it is important to maintain the correctness of labels with respect to the newly produced data. The reference annotations are transformed according to the augmentation method. Figure \ref{fig:data-augmentation} illustrates time stretching and block mixing \cite{parascandalo2016} applied to an input signal and its corresponding annotation. The illustration uses the spectrogram for representing the audio signal, for better visibility of their effect on the signal content.
In time stretching, the duration of the audio is extended, and the activations of sound classes are extended accordingly, to account for the labels being time-stretched too. In block mixing, two different blocks of the same audio recording are additively mixed, and the activations of sound classes are combined using an AND operation, to account for the sounds being active in the resulting mixture when they are active in at least one of the original blocks. 
In contrast, in mixup data augmentation the original data blocks and their labels are combined as a weighted sum, resulting in new input data for which the labels are no longer binary.

\subsection{Feature representation}

The most commonly used features in sound event detection are currently log mel energies \cite{Mesaros2018_TASLP}, which represent the audio signal energy using perceptually-motivated frequency and amplitude scales. 
Processing and representations inspired by the human hearing have been found to be useful for a wide range of audio classification tasks, including automatic speech recognition, music information retrieval, sound event detection, and acoustic scene classification.

To obtain the mel spectrum, the spectrogram of an audio signal is redistributed on the mel scale using a filterbank. 
The mel scale is a perceptual scale of pitches judged by listeners to be equal in distance from one another. The scale was obtained experimentally, with listeners providing a set of points that relate the real frequency with the perceived pitch. 
The reference point between mel and Hz is defined by assigning a perceptual pitch of 1000 mels to a 1000 Hz tone, resulting in a linear relationship between the two scales under this point, and approximately logarithmic over it. 
A few approximations exist for the mapping, of which the most commonly used is: 
\begin{equation}
    mel(f)=\frac{1000}{\log2}\log \left(1+\frac{f}{1000}\right)
\end{equation}
The process is implemented using triangular filters equally spaced on the mel scale. The energy coefficients in each band are obtained as a weighted sum of the spectral amplitude components falling within the filter range, with the weights given by the filter amplitude values \cite{serizel2018acoustic}.
The number of filters varies. Early SED systems used 40 filters, motivated by the good performance of the setup in automatic speech recognition. In automatic speech recognition, the number of filters is chosen to be sufficient for representing the coarse shape of the spectrum that allows different phonemes to be identified. However in SED, using more filters is often beneficial, for example, because it allows resolving overlapping sounds at closely spaced frequencies. As a result of the advancement of machine learning techniques that allow learning acoustic models from features of higher dimensionality, recent SED studies have used up to 128 filters.
Most commonly, the signal energy is measured on a logarithmic scale, resulting in the log mel energy representation.

In previous work in the speech and music domain, the next processing step was the discrete cosine transform (DCT), resulting in mel-frequency cepstral coefficients (MFCC). 
DCT transforms the spectrum into a cepstrum, in which the lower order coefficients represent the coarse spectral characteristics and higher order ones are related to fine details. In addition, DCT has strong compaction properties, resulting in most of the signal energy being concentrated in the lower order components of the output, which allows truncation of the feature vector to the lower order coefficients without losing much spectral information. DCT provides both compression and decorrelation of the information in the signal spectrum. 
Decorrelated features were important for classical methods using GMMs, as they allowed use of diagonal instead of full covariance matrices, therefore fewer parameters for the model. 
The overall feature density modeled using a set of full covariance Gaussians can also be obtained by using a large enough set of diagonal covariance Gaussians \cite{Reynolds2015}, and in consequence the choice of model configuration was often dictated by the amount of data available for estimating the parameters. While full covariance GMMs are able to fit the data best, diagonal covariance GMMs are a good compromise between quality and model size, allowing robust parameter estimation from small amounts of data, which was an important design factor when computation power and training dataset sizes were limited.
MFCCs were the dominant feature of choice in the early environmental sound classification studies \cite{barchiesi2015acoustic}.

With the shift to deep learning and the increase in computational resources and datasets, best accuracies are typically obtained with large DNN models. When SED algorithms are implemented in resource-constrained devices, the model size is 
a trade-off between accuracy and computational complexity. Because DNNs are capable of learning from correlated features, the decorrelation introduced by DCT is no longer necessary.  
Even though neural networks are able to learn features directly from raw audio input, engineered time-frequency representations are still more common in comparison to end-to-end approaches, since end-to-end approaches would require substantially larger training datasets, which are not that easily available for SED.

A recommended feature representation for a basic SED system is 40 log mel energies calculated in 40 ms windows, which provide a detailed enough, but not too high-dimensional representation. The window length is a compromise for the large variety of sounds, some that may be more harmonic (suitable for analysis using a long window), and others that may be short, transient-like (requiring  a short analysis window). This is expected to produce reasonably good performance, without requiring excessive training time. 

Besides the spectrogram and mel spectrogram, there are other possible time-frequency representations. The spectrum can be decomposed into the harmonic and percussive (noise-like) components \cite{fitzgerald2010harmonic} and separate models can be trained on the two spectra, under the assumption that sound events of different types will be modeled better by either component, especially when these sounds  overlap.
Another time-frequency representation that can be used for SED is the constant-Q transform (CQT), in which the frequency axis is logarithmic \cite{serizel2018acoustic}. CQT offers better spectral resolution at lower frequencies and better temporal resolution at higher frequencies. Multiple spectral representations can also be used, for example spectrograms calculated at different analysis resolutions can also be used \cite{Adavanne2017}. The use of representations with different time or frequency resolutions is motivated by the assumption that different types of sounds may benefit from modeling at different frequency or time resolution, e.g., short sounds, such as a door slam benefit from high time resolution in analysis, while the harmonic components of sounds such as birds singing may be better represented using a high resolution in frequency.

A recently studied alternative to the above so-called handcrafted features is \textit{feature learning}. Unsupervised feature learning methods attempt to learn time-frequency representations directly from the data, without using expert knowledge on specific attributes of the data, as is the case with the handcrafted features. Approaches using non-negative matrix factorization (NMF) applied to CQT \cite{Bisot2017} and spherical k-means applied to mel spectrograms \cite{Salamon2015} were successfully used for acoustic scene and sound classification. On the other hand, for polyphonic sound event detection, attempts to learn filterbanks and equivalent time-frequency representations directly from the raw waveform did not outperform the equivalent system based on the handcrafted alternative \cite{Cakir2018}. 

Feature learning may also be incorporated into the overall machine learning process. As with deep learning in general, feature learning requires large amounts of training data when raw audio is used as its input. This is evident from the feature learning examples given above: when applied on time-frequency representations that have a reduced dimensionality in comparison to the original audio, the learned features provided state-of-the art performance, but learning from the raw audio did not bring any advantages. In comparison, domains which apply feature learning more successfully, like automatic speech recognition, use large amounts of data, which allows more robust learning of the necessary transformations. 
Furthermore, the convolutional neural networks as used for sound event detection and classification learn higher-level feature representations using the input spectrograms, so the convolutional components are essentially doing feature learning without being explicitly called so, while the last layers handle the classification decision.

\section{Machine learning for sound event detection} 

Deep neural networks have brought tremendous improvement in many domains such as image classification and speech recognition, and are now also the dominant approach in environmental sound analysis and classification, as observed in the recent years \cite{Mesaros2018_TASLP, Mesaros2019_TASLP}. Their main drawback is that they require large amounts of data for training. This need for large datasets is a problem for sound event detection because the domain still lacks large datasets of strongly-labeled data. Advanced training strategies involving weak labels and transfer learning are providing suitable solutions to cope with shortcomings in the data, but the general system architectures often do not change dramatically.

\subsection{Convolutional Recurrent Neural Networks}

\begin{figure*}
    \centering
    \includegraphics[width=0.90\textwidth]{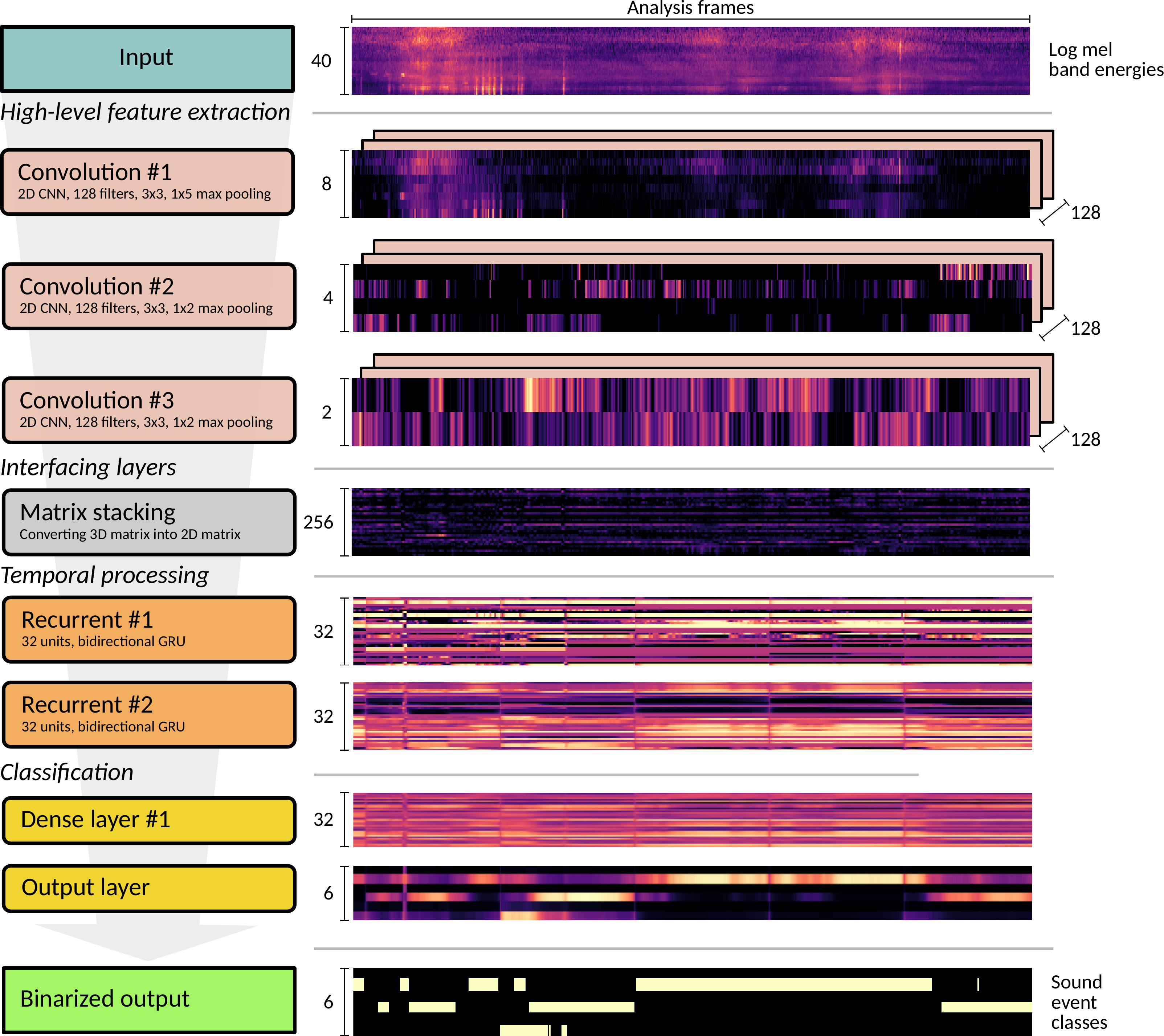}
    \caption{CRNN architecture example and corresponding processing of the feature representations through its layers.}
    \label{fig:crnn-basic}
\end{figure*}

A general-purpose network architecture for sound event detection is the convolutional recurrent neural network (CRNN), containing convolutional and recurrent layers that have specific roles \cite{Cakir2017_TASLP}. 
The convolutional layers act as feature extractors, aiming to learn discriminative features through the consecutive convolutions and non-linear transformations applied to the time-frequency representation presented at the input of the network. The recurrent layers have the role of learning  the temporal dependencies in the sequence of features presented at their input.
Figure \ref{fig:crnn-basic} presents a CRNN architecture consisting of three convolutional blocks, followed by two recurrent layers and two feed-forward layers. The processing of the information and resulting representation after each main component of the network is presented in the figure along with the structure. The network receives as input a time-frequency representation of the data, in this case log mel energies calculated using 40 filters for a segment of data of length $T$, and outputs event activity probabilities for the target sound event classes.  

A convolutional block consists of a convolutional layer, a nonlinearity, and pooling. The convolutional layer applies a kernel to its input to create a new feature representation. This representation goes through a nonlinear activation layer, usually rectified linear units (ReLU). The rectified linear function, defined as $g(z)=max\{0,z\}$, is a piecewise linear function that provides a transformation to a nonlinear feature space. The piecewise linear function preserves many properties of linear models, allowing optimization using gradient-based methods \cite{Goodfellow-et-al-2016}. Pooling and subsampling are then used to reduce the dimensionality of the data. 
In the example in Fig.~\ref{fig:crnn-basic},  the first convolutional block reduces the number of frequency bins to 8 through a 1x8 max pooling operation applied to the outputs of its 128 filters. Because sound event detection requires estimation of the temporal position of sound events, the time axis must be maintained, and therefore the pooling operation is performed on the frequency axis only. The next convolutional block reduces the frequency bins to 4 (1x2 max pooling), and the last one reduces it to 2 (1x2 max pooling). Repeated convolution blocks consisting of the convolution, nonlinear activation and pooling operations allow consecutive transformations of the input, illustrated in panels 2--4 of Fig.~\ref{fig:crnn-basic}, corresponding to the convolutional blocks. These are features learned by the network from the data, even though for the human eye they might not represent any meaningful pattern.

A recurrent layer has the role of learning contextual information from the evolution of the signal over time. Based on the successive time segments provided to its input, the recurrent layer receives temporal context information for how the sound events start, end, and follow each other, and the internal temporal structure of the events themselves. Some specialized recurrent units are bidirectional, which means they take into account both past and future inputs, in addition to the current one. By including information from previous (and possibly future) inputs into the current output, a recurrent layer is assumed to provide predictions that take into account the length of event instances as seen in the training data. In practice, the temporal context is limited and its contribution to the output is determined by the learned weights. In the given example system \cite{Cakir2017_TASLP}, the recurrent layers are implemented using gated recurrent units (GRU), having as inputs the output of the current frame of its preceding layer, and the output of the previous frame of the current layer. In addition to GRUs, long short-term memory (LSTM) layers are a commonly used recurrent network structure.

The outputs of the last convolutional block are stacked over the frequency axis, yielding a 2D feature representation which is then provided as input to the first recurrent layer. In the illustrated example, after the last convolutional block the feature representation has been reduced through pooling to 2x128xT (indexed as frequency, filter number, time); the 128 filter outputs are stacked to form a 256xT feature matrix, which is then processed by the recurrent layer with 32 input units.

The feedforward layers have the role of producing sound event activity probabilities based on the output from the last recurrent layer. There can be a number of feedforward layers that use sigmoid units before the output layer, which is selected according to the task (e.g., softmax for classification). In this case, the output layer consists of sigmoid units, and these provide probabilities individually for each target event class.

Because the expected output in SED is a sequence of binary activation indicators for each event class, the network output is binarized. The threshold for binarization can be selected simply as 0.5, or can be optimized for the problem based on statistics of the data. Furthermore, the resulting binary sequences can be postprocessed to form event instances, and can use heuristic rules based on expected/plausible sound event lengths, and unlikely activation patterns, as explained in Section \ref{sec:general-ml}.

The entire network architecture is trained using the backpropagation algorithm, which allows computing and propagating information through the network from the output towards the input. Learning uses gradient-based optimization and a cost function with respect to the network parameters that measures the error between the predicted output and the target output \cite[p.198]{Goodfellow-et-al-2016}. 
In sound event detection, the network training aims to minimize the cross-entropy--a measure of two probability distributions diverging from one to another--between the true labels and predicted labels \cite{mcfee2018statistical}. Each training example has a probability of 1.0 for the correct class, and a probability of 0.0 for all other classes, while the network estimates the probability of an example belonging to each class, and cross-entropy is calculated between the two.

In a practical design process, the network goes through incremental changes in adjusting the architecture and hyperparameters, until a best performance is achieved. 
The hyperparameters are variables that determine the network structure (e.g., the number of hidden units in each layer, dropout) and the variables which determine how the network is trained (e.g., learning rate). Hyperparameters influence the optimization process, and finding a balance between training error, generalization properties, and computational resources is a problem itself.
A higher number of hidden units increases the model capacity to learn complex relationships, but increases the number of parameters and therefore requires more training data and computational resources, and may lead to overfitting (learning too well the training data, detrimental to generalization when tested on new data). On the other hand, a smaller number of hidden units may lead to underfitting. Increasing the convolutional kernel size also increases the model ability to learn complex relationships in the training data, but increases also the number of parameters to be trained. 

Dropout is a regularization technique used with large networks to avoid overfitting by randomly dropping out (ignoring) a number of units in the network during training. In effect, a layer with dropped out units looks like a layer with a different number of units and connectivity to the prior layer, so each update is performed with a slightly different “view” of the layer. This in turn makes the model more robust to small changes. An often used value for dropout is a probability of 0.5 for retaining the output of each node in a hidden layer, but values of 0.7 - 0.8 are also commonly encountered. 

Probably the most important hyperparameter is the learning rate, which defines how quickly a network updates its parameters. The learning rate determines the behavior of the gradient descent: a low learning rate slows down the learning process but converges smoothly, while a large learning rate speeds up the learning but may not converge. For this reason, typically a decaying learning rate is used, which starts with large values and is gradually reduced as the training progresses.
Other hyperparameters include the number of epochs--the number of times the whole training data is shown to the network while training, which should be sufficiently large--and batch size--the number of sub samples given to the network after which parameter update happens, for which a typical default value is 32. 

Hyperparameter selection can be done by manual search, or as a grid search, in which a network is trained and tested for all combinations of hyperparameters, leading to a very high computational cost for finding the best solution. An efficient technique for hyperparameter selection is the randomized search, in which random combinations of the hyperparameters (instead of all combinations like in grid search) are used. A detailed discussion on the practical methodology in hyperparameter selection for deep learning is presented by Goodfellow et.al. \cite [Chapter 11] {Goodfellow-et-al-2016}. 

The network architecture is selected according to the task: recurrent layers are included when sequences of data are modeled and predicted, while in classification tasks there is no need to preserve temporal information so the networks usually include only convolutional blocks
\cite[p.415]{Goodfellow-et-al-2016}. 
The choice of architectures for a network doing sound event detection is commonly based on architectures shown to perform well in similar audio classification problems. The size of the networks is often limited by the availability and form of the training data in terms of the number of classes and training examples. 
Architectures with 2--5 convolutional blocks are most commonly encountered \cite{Mesaros2019_TASLP}, while the number of recurrent and fully-connected layers that follow is usually 1--2 each. The optimal numbers and sizes of the convolutional filters and the size of the input time-frequency representation are generally chosen through multiple trials using a validation dataset--a separate dataset that consists of data not seen during training and acts as test data for verifying the system performance. The size of the network is influenced by the complexity of the task and the amount of available data for robustly learning the model parameters. This influences both the number of layers and their sizes. Too deep networks will overfit a small dataset, while too shallow networks will not offer the benefit of modeling complex enough relationships between the inputs and outputs. In comparison, in speech recognition tasks, where hundreds of hours of data is available for training,  networks with tens of layers are encountered, while in acoustic scene classification, 7-9 convolutional layers are common with datasets of 20-40 hours.

\subsection{Advanced methods} 

Transfer learning offers a different solution to the data scarcity problem, as an alternative to data augmentation. The main idea is to take advantage of large amounts of data which may be available for certain tasks, and use it to solve a target task. A neural network is therefore trained to solve a pretext task (a predesigned task that allows learning of acoustic features) and then the pretrained weights are used to construct the network for the target task \cite{Cramer2019, Jung2019}. The representations provided by the pretrained layers are called \textit{embeddings}, and can be considered as input features for a downstream task (the original target task, a task in which the learned audio features will be used to solve). Available examples of pretrained networks that can be used to calculate embeddings include VGG-ish \cite{hershey2017cnn}, SoundNet \cite{aytar2016soundnet}, and $L^3$-Net \cite{arandjelovic2017look, Cramer2019}.

For example, $L^3$-Net learns embeddings using the pretext task of audio-video correspondence between one video image frame and one 1-second audio segment. The network structure contains two branches, one for audio and one for video, with the trained audio branch being later used to produce the embeddings. Each branch consists of four convolutional blocks, the outputs of which are flattened (converted into a 1-dimensional array) and concatenated, then fed into fully-connected layers that output the audio-visual correspondence probability. $L^3$-Net \cite{Cramer2019} offers design choices regarding input representation and training data domain, and was shown to outperform SoundNet and VGGish in the downstream task of environmental sound classification. 

Another solution for dealing with the scarcity of data is to use weak or noisy labels in training. While weak labels refer to labels that do not provide temporal information, noisy labels are labels that may be incorrectly attributed, such as automatically generated \cite{Gemmeke2017} or non-curated labels \cite{fonseca2019learning}. 
With weak labels, the key challenge for the system is coping with a more challenging learning process in which it must learn the temporal position of the labeled sound without supervision. One common approach for weakly-supervised learning is \textit{multiple instance learning}, in which the frames of the signal are considered to be training instances that are presented as bags. A bag is an entire weakly-labeled example consisting of multiple frames (instances), and the weak labels are therefore attached to bags rather than to the instances in each bag. Bags are negative and positive examples for the target classes: negative bags contain only negative instances while positive bags can contain both positive and negative instances \cite{kumar2016weakly}. In learning, the neural network predicts class probabilities at instance level, while the pooling functions aggregate instance-level information to bag-level, to be used for minimizing the loss at bag-level \cite{kumar2016weakly}. Other solutions for weakly supervised training include \textit{attention-based neural networks}, which contain a module that predicts the importance of the frames in the signal\cite{xu2017attention}.
The attention mechanism has the role of helping the model distinguish between relevant and irrelevant parts of the audio clip in the specific SED task addressed. It is conceptually similar to multi-instance learning, in the sense of using instance-level predictions that are aggregated into a bag-level prediction. The bag-level prediction is a weighted sum of the instance-level predictions, with the weights essentially being the attention function that determines the selection of the important frames for a target class. The attention mechanism is usually implemented as a layer of the neural network model, and the weights are learned during training.

Large amounts of user-generated audio material are available as web audio for which labels can be inferred from user-generated metadata. Such data is referred to as having noisy labels because there is no guarantee that the labels are correct. For example AudioSet \cite{Gemmeke2017} consists of 5000 h of audio from YouTube which was labeled automatically and then partially verified, and has an estimated label error of above 50\% for 30\% of the  classes\footnote{See https://research.google.com/audioset/dataset/index.html for an explanation of the quality assessment. Information accessed December 2020}. The effect of noisy labels manifests as increased complexity of learned models, and often decreased performance. Noisy labels can be handled using for example iterative verification loops that fine-tune the models based on prediction consensus and the original labels \cite{Dorfer2018}, or by using noise-robust loss functions which rely more and more on the model predictions than on the original noisy labels as learning progresses \cite{fonseca2019learning}.

Another category of advanced methods used in machine learning are student-teacher methods \cite{tarvainen2017mean}, a form of knowledge distillation that transfers knowledge from one network to another. Knowledge distillation can compress a large model (the teacher) by teaching a smaller one (the student) to mimic the behavior of the teacher by replicating its outputs. In student-teacher methods, the student is trained with the outputs of the teacher, instead of the reference target outputs, as a form of soft target distribution over classes \cite{hinton2015distill}.

Student-teacher models are encountered in the problem of weakly-labeled semi-supervised sound event detection. The task is viewed as two separate subtasks: one of audio tagging and another of boundary detection. One presented solution was to train a teacher model on a coarse resolution in order to perform well in audio tagging, and guide the student model to learn boundary detection at a finer resolution, using unlabeled data \cite{Lin2020}. Another approach employed student and teacher networks each having two branches: a coarse temporal resolution branch designed for audio tagging, and a fine-resolution one designed for the detection task \cite{Yan2020}, with each branch in the teacher model teaching the corresponding branch in the student model. 

\section{Performance evaluation}

Ideally, the performance of a sound event detection system should be measured by user satisfaction in the application where it is used \cite{krstulovic2018audio}, in order to answer the user expectations of quality. During development, simpler computational metrics are used to assess performance by comparing the system output to reference annotation. The comparison provides statistics on the correct and erroneous detection and calculates common metrics used in pattern classification. 

The system outputs a decision for each consecutive audio segment and class, and this output is compared to the reference annotation to determine the system performance. 
Even though the output time resolution of many SED systems is similar to the analysis resolution (tens of milliseconds), doing the comparison at this resolution may penalize small timing errors which do not matter in practical applications.
Furthermore, if the ground truth is produced by human annotators, the timing of the ground truth is subjective \cite{mesaros2018datasets}, which makes the comparison at too high temporal resolution unreliable. One approach is to compare the system output and reference annotation on a fixed temporal grid which is more coarse than the system output resolution, for example one second. This is referred to as \textit{segment-based evaluation}. 
On the other hand, \textit{event-based evaluation} compares the system output and the reference annotation in terms of sound instances, checking how well the onset/offset times of the detected sound instances correspond to the times of the  annotated sound instances, event by event. 

\begin{figure*}
    \centering
    \includegraphics[width=0.9\textwidth]{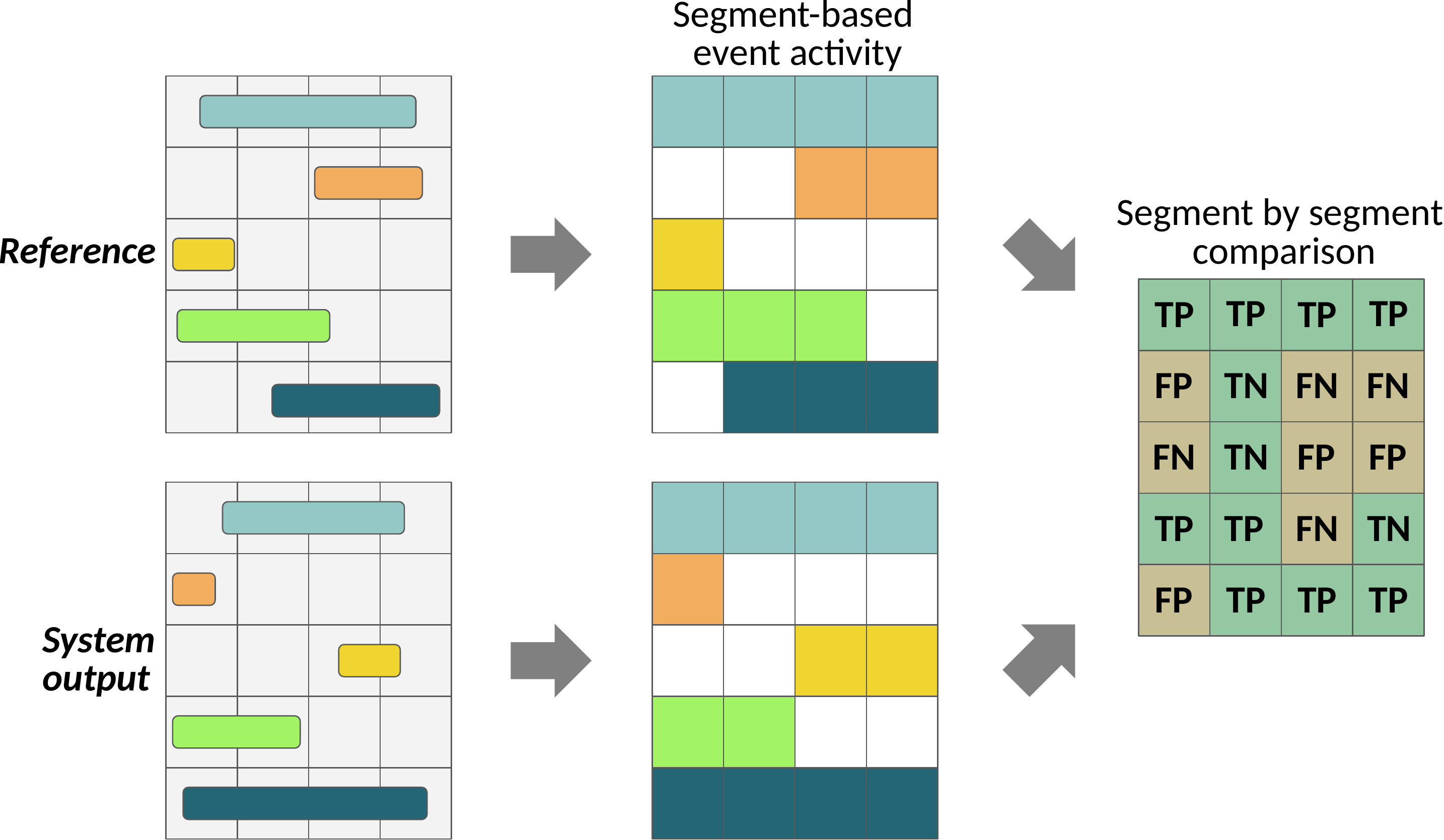}
    \caption{Segment-based evaluation: comparison is made at segment resolution, which quantizes both the system output and the reference annotation temporally. Activity of sounds events in each evaluation segment is marked with a binary indicator and these segment-based activities are then compared.}
    \label{fig:segm_eval}
\end{figure*}

The segment-based evaluation process is illustrated in 
Fig.~\ref{fig:segm_eval}. The evaluation process will quantize sound event onsets and offsets to the used evaluation grid, extending the sound event length] such that the activity indicators cover all segments in which the sound is active, even for a very short time. The comparison between the reference annotation and system output uses the segment-level binary activity indicators to count the correctly and erroneously detected events in terms of \textit{true positives} (TP), \textit{true negatives} (TN), \textit{false positives} (FP), and \textit{false negatives} (FN), which can be used to calculate different measures of performance. 

Metrics commonly used in pattern classification include precision (P), recall (R), F-score and error rate (ER). They are defined as based on the counts of correct and erroneous detections as follows: 
\begin{equation}
    P = \frac{TP}{TP+FP},
\end{equation}

\begin{equation}
    R = \frac{TP}{TP+FN},
    \label{eq:R}
\end{equation}   

\begin{equation}
    F = \frac{2PR}{P+R}
\end{equation}
While P, R, and F-score regard and count the errors individually, ER counts a joint occurrence of a false positive and a false negative as a single \textit{substitution} error \textit{S}. This procedure originates from speech recognition evaluation and is based on the \textit{edit distance}, which quantifies how dissimilar two strings (e.g., words) are to one another by counting the minimum number of operations required to transform one string into the other. 
False positives unaccounted for in \textit{S} are the remaining \textit{insertions} \textit{I}, and false negatives unaccounted for in \textit{S} are the remaining \textit{deletions} \textit{D}. ER is calculated as the total number of errors with respect to the number of reference events \textit{N}:
\begin{equation}
    ER=\frac{S+D+I}{N}. 
\end{equation}
The main drawback with these metrics is the trade-off between correct and missed detection, requiring selection of the operating point (the trade-off point) according to the target application and the cost of the different error types. 
On the other hand, metrics based on receiver operating characteristic (ROC) curve characterize performance at a range of operating points by plotting at various threshold settings the true positive rate (TPR, also called sensitivity, or recall), calculated according to \eqref{eq:R}, against the false positive rate (FPR) defined as:
\begin{equation}
FPR = \frac{FP}{FP+TN}.
\end{equation} 
A single metric describing this curve is the area under the curve (AUC), which is equal to the probability that a classifier will rank a randomly chosen positive instance higher than a randomly chosen negative one \cite{mesaros2018datasets}. 

\begin{figure*}
    \centering
     \includegraphics[width=0.9\textwidth]{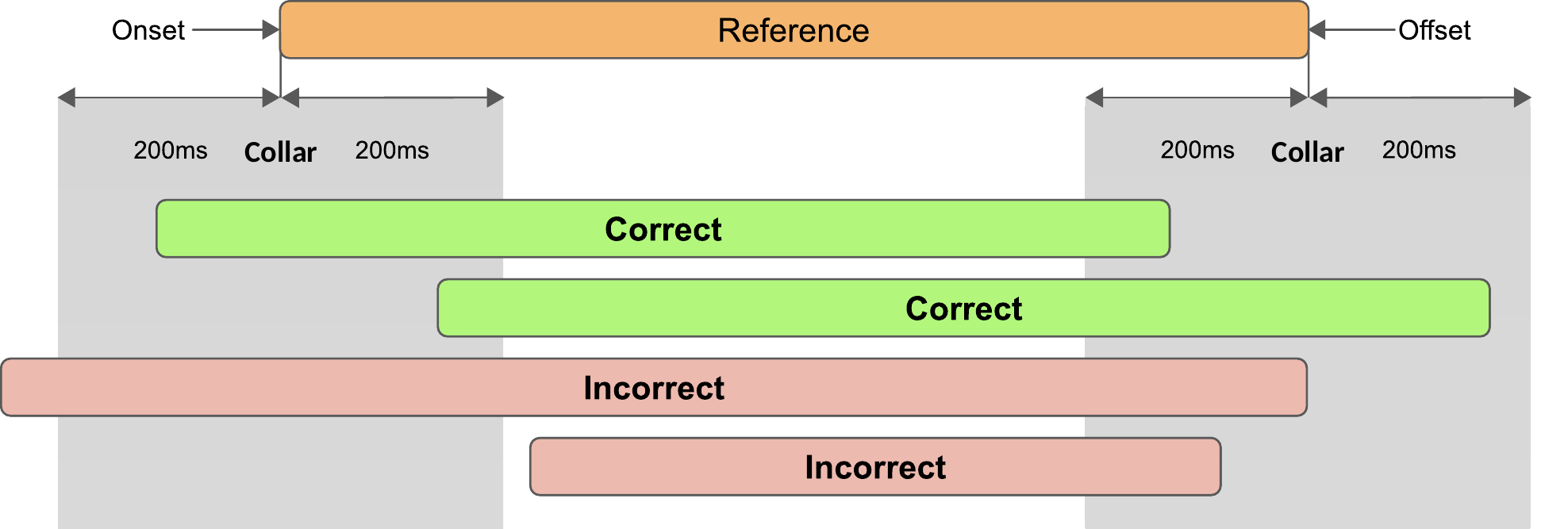}
    \caption{Event-based evaluation with a 200 ms collar on both onset and offset.}
    \label{fig:ev_based_tolerance}
\end{figure*}

For event-based evaluation, the subjectivity of the temporal boundaries is alleviated by allowing some degree of misalignment called a \textit{collar} between the compared events in the reference and system output. 
An event instance is counted as a true positive if it has the same label as the corresponding reference event, and its temporal boundaries lie within the permitted temporal collar with respect to the reference event. The condition can be used only for onset, or for both onset and offset.  Figure \ref{fig:ev_based_tolerance} illustrates the possible cases when both onset and offset condition are used with a 200 ms collar. 
Event-based evaluation is more intuitive for humans interpreting the result of a sound event detection system, because it expresses the performance in terms of sound instances being detected. 

The segment-based and event-based evaluation options represent two conceptually different views of the same output: segment-based evaluation represents the performance of the system in detecting the temporal regions where sound events are active, while event-based evaluation represents the performance of the system in detecting individual instances of sound events. This is similar to metrics used in multi-pitch transcription, which can be calculated at frame level or note level. However, multiple pitch estimation (frame-level information) and note tracking (event-level information) are seen as separate tasks: the pitch estimation task requires reporting pitch values that are active in every frame, while the note tracking task requires reporting a list of notes, where each note is designated by its F0, onset and offset times \cite{bay2009}. Sound event detection defines the task at event level, but practical methods solve it at segment level, thus creating a mismatch between the problem and its solution. The choice of metric is therefore left to answer the requirements of the application rather than the task definition.

The use of collars is a significant weakness in event-based evaluation, and while the segment-based evaluation alleviates to some extent the subjectivity arising from the temporal boundaries, the field is still open for development of more suitable metrics. One recent proposal, the polyphonic sound detection score (PSDS), redefines the definition of true and false positives based on intersection between the system output and reference, making the evaluation tolerant to segmentation of event instances \cite{Bilen2020}. PSDS is defined as the normalized area under the ROC curve, therefore also counteracts the dependence on operating point of the most common metrics.

Since sound event detection has several applications, each of them dealing with different types of data and different performance requirements, it has been important to establish common evaluation protocols in order to allow comparing different methods. The international evaluation challenge on Detection and Classification of Acoustic Scenes and Events (DCASE) is an important community effort that offers annually several tasks allowing public benchmarking of methods. DCASE Challenge provides open data in order to allow reproducibility of results, and provides public evaluation of systems submitted by participants, with complete results and technical descriptions of submissions made available on the challenge website\footnote{\url{dcase.community}}.

\section{Related research problems}

Many scientists and engineers starting work on sound event detection may already be familiar with related audio research topics. Sound event detection shares many similarities to other audio analysis tasks studied in neighboring fields such as speech and music analysis, but there are also many fundamental differences.

The biggest difference between sound event detection and several classification tasks such as acoustic scene classification, music genre classification, and speaker recognition is that classification characterizes one entire audio recording with one class label, whereas sound event detection aims at estimating temporal activity of sound classes within the audio recording. This requires machine learning architectures that produce classification output in consecutive segments, contrary to a single classification output for the entire duration of the test audio.
Despite this difference, machine learning techniques used for classification can be adapted to sound event detection by making some changes to the classifier architecture, typically omitting temporal pooling operations from the structure of the neural networks so that the systems produce output at multiple time points. 

A significant difference to various speech analysis tasks such as automatic speech recognition is that they typically target one source (speaker), whereas sound event detection in general is targeted towards detecting multiple overlapping sources. Therefore, sound event detection systems need to produce polyphonic output. This is similar to some music analysis tasks, such as automatic music transcription or multiple instrument recognition, since in music there are often multiple notes or instruments playing simultaneously. As discussed earlier, polyphonic output can be obtained by multi-label classifiers. For example, a neural network based system can be used to produce multi-label output by having sigmoid nonlinearities for the output nodes.

Most audio augmentation methods used in other  domains are applicable for sound event detection.
What makes sound event detection different from some other fields is the existence of many different combinations of overlapping sounds. Therefore, mixing segments of existing data and its annotations can be used in SED but not necessarily in other audio tasks.

In terms of acoustic features, sound event detection solutions comprise techniques that are similar to other audio analysis fields. Target sound events can have quite diverse acoustic characteristics, including transient, harmonic, and noise-like components. Speech signals also contain all these components, therefore log mel energies originating from speech analysis can be directly applied to SED. 
In comparison to speech and music analysis fields, the amount of data available for SED research is still rather limited. For speech and music research, several large-scale datasets have been produced \cite{Baumann2018, Bertin-Mahieux2011} and successfully used in benchmarking systems. However, there is no established taxonomy of sound events, and therefore joint community efforts for collecting large-scale SED datasets have been much more limited. Each application has slightly different sound classes, annotation requirements, and is used in different environments. In addition, labeling of sounds is subjective in the selection of terms used to describe sounds \cite{guastavino2018everyday}, which introduces the additional burden of processing a diverse set of labels or imposing a strict vocabulary in the data collection process.
Some notable audio data resources are AudioSet \cite{Gemmeke2017} and FSDnoisy18K \cite{fonseca2019learning}, but they are weakly labeled and contain label noise, which affects their use. 

The focus of this paper has been on single-channel methods, but the use of multiple microphones can potentially improve the detection accuracy and also allow joint detection and localization of sound events. For processing multichannel audio, techniques used in other fields can be borrowed, for example enhancement of the target signal, such as in distant automatic speech recognition. 
Multichannel SED systems (or joint localization and detection systems) use  either intra-channel features that represent spatial properties of the input audio, or calculate low-level features separately from each channel, and then use machine learning models that enable learning the spatial properties. The spatial features include those commonly used in spatial audio processing, such as inter-channel cross-correlation and time-difference \cite{Adavanne2017}. Low-level features include typically the channel-wise magnitude or phase spectra as the input. The use of CNNs enables learning spatial features from these, such as as level or phase difference between channels.

One notable difference between joint detection and localization and the more traditional localization studies is the need to associate estimated sound classes to their estimated directions. Traditionally, localization studies targeted only a single predefined sound class, most often speech, even though some cases deal with multiple speakers. During training and inference, joint detection and localization must have in place specific procedures to associate a certain sound class to a certain estimated location  \cite{Trowitzsch2020}. Localization can naturally be extended to moving targets \cite{politis2020dcase}, leading to joint detection and tracking. 

\section{Future perspectives}

The capability of deep learning techniques to learn complex models from large datasets has meant that the lack of large annotated suitable datasets is often the bottleneck in research and development. Instead of the traditional approach where an annotated dataset is first collected and then models are trained based on supervised learning, researchers have started to look into alternative ways for obtaining acoustic models. 

\emph{Active learning} refers to methods where a learning algorithm selects an unlabeled audio sample from a dataset and asks an annotator to label it. The selection is done in such a way that the model accuracy is improved the most. Optimal selection of the samples to be annotated will result in many fewer annotated samples for learning robust models. For example, Zhao et al. \cite{Zhao2020active} found that an active learning system requires only 2\% of the annotations needed by the standard supervised learning algorithms, in order to reach the same accuracy. This kind of human-in-the-loop approach can involve feedback from end users for improving acoustic models.

When audio from end users is used for training models, it creates concerns on how the privacy of the data is handled, since audio contains personal information such as conversation.
Recent efforts for privacy-preserving techniques concentrate around \textit{federated learning} \cite{mcmahan2017communication}, as a form of collaborative machine learning that extracts information from multiple users' data without transferring or storing any of this data in the cloud. Federated learning aims to learn or improve a general model using local data, then transfer only the model information. Information originating from multiple users is averaged to improve the shared model, and no individual model updates are stored \cite{konevcny2016federated}. Current application examples include image classification and language modeling tasks. However, they require both special machine learning techniques that can operate in a distributed way, and a suitable infrastructure for  communication and secure aggregation of the information. The development of suitable techniques for wide deployment of federated learning is still in its infancy, but it is reasonable to expect need of such techniques in sound event detection too, in a future likely to adopt use of smart home surveillance systems and intelligent personal assistants. 

An alternative to collecting new audio is to develop methods for transferring knowledge from different domains to recognize new classes of sounds. One emerging transfer learning technique that does not require any audio samples from new classes, \emph{zero-shot learning}, uses auxiliary information about the new classes such as their textual descriptions in order to obtain an acoustic model \cite{xie2019zero}. A zero-shot learning stage includes data from a large number of training classes and auxiliary information about them, such as their textual descriptions of attributes. During training, a zero-shot model is trained to model the relationship between acoustic features and the auxiliary information. At usage time, the zero-shot model is used to do recognition of new classes by modeling the relationship between audio samples and the auxiliary information from the new classes. Until now, zero-shot learning has been used for classification without estimating temporal activities of classes \cite{xie2019zero}, but the temporal classification methods presented in this paper allow extending the method to SED. 

In addition, similarly to other fields, current deep learning solutions for sound event detection are sensitive to mismatches between the training and test data. In sound event detection these kind of mismatches are present easily, for example when the methods are deployed to different devices. To improve the general robustness of sound event detection methods, one solution is \emph{model adaptation}, which changes the model parameters to retain the performance in new conditions. The adaptation can be done, for example, in supervised manner, by retraining an existing network with training data available from the new environment, or in unsupervised manner, by changing the feature extraction so that the statistics of the features match between the training and test data. 

The robustness and other relevant tasks have recently been addressed in the DCASE public evaluation campaign. As the research in the field progresses, new methods for benchmarking should be introduced to address arising limitations of the technologies.

\bibliographystyle{IEEEtran}
\bibliography{references.bib}

\clearpage

%\begin{mdframed}[backgroundcolor=lightgray!50] 
\section*{Appendix: Neural networks for audio classification} 

\noindent A \textbf{feed-forward neural network} is a machine learning model defining function $\hat{y}=f(\mathbf{x},\mathbf{\theta})$ that maps input vector $\mathbf{x}$ to output $\hat{y}$.
In supervised training of the network, parameters $\mathbf{\theta}$ are estimated based on a set of example training pairs ($\mathbf{x},\hat{y}$).

A neural network is composed of \textit{layers}, which are composed of \textit{units (neurons)} working in parallel. The number of layers between the input and the output gives the depth of the model.
Each unit receives an input from units of the previous layer and computes its own output based on an \textit{activation function}.
The activation functions used are typically nonlinear, in order to allow modeling of complex, nonlinear relationships between inputs $\mathbf{x}$ and output $\hat{y}$.
The training examples specify what the output layer should do for each data point $\mathbf{x}$, but do not provide a desired output for the layers in between the input and output. For this reason, these layers are called \textit{hidden layers}, containing \textit{hidden units}. The input $\mathbf{x}$ propagates through the layers and produces an output $\hat{y}$, in what is called \textit{forward propagation}. 

Training a neural network uses gradient-based learning, which aims to minimize a cost function defined to measure the error between the predicted and the expected output. The function $f$ mapping the inputs to outputs is usually highly complex due to the consecutive layers composed of non-linear units, making gradient computation difficult. 
The \textit{back-propagation algorithm} is a method that allows information to flow backward through the network in order to compute the gradient, for training the network using iterative gradient-based optimization.

A simple binary classification problem requires the prediction of the value of a binary variable $y \in \{0,1\}$. In this case, the network output is typically interpreted as the probability $f(\mathbf{x},\mathbf{\theta}) = P(y=1|\mathbf{x})$, $P\in[0,1]$.
This output is obtained with the use of a sigmoid unit defined by
\begin{equation}
    \hat{y} = \sigma \left(\bm{w}^T \bm{h}+b \right), \end{equation}
where $\sigma$ is the logistic sigmoid function $\sigma(x)= \frac{1}{1+\mathrm{exp}(-x)}$, and weight vector $\bm{w}$ and scalar bias $b$ define the model $\theta$ that is applied to the input vector $\bm{h}$ of the unit. The output calculation can be seen as being composed of two parts: a linear layer that computes $z=\bm{w}^T \bm{h}+b$, and a sigmoid activation function that converts $z$ into a probability. For the complete explanation, we refer the reader to \cite [p.177] {Goodfellow-et-al-2016}.

\textbf{Multi-class single label classification} tasks require the output of the network to represent a probability distribution over $n$ different classes. In this case, the output is a vector $\hat{\bm{y}}$ with $\hat{y}_i=P(y=i|\bm{x})$, for which each element is between 0 and 1 (i.e. a probability) and the entire vector sums up to 1 (i.e. it represents a valid probability distribution). A linear layer consisting of weight matrix $\mathbf{W}$ and bias vector $\mathbf{b}$ computes vector  $\bm{z}=\mathbf{W}^T \bm{h}+\bm{b}$,  followed by the softmax function which exponentiates and normalizes $\bm{z}$ \cite [p.179] {Goodfellow-et-al-2016}:
\begin{equation}
    \mathrm{softmax}(z)_i = \frac{\mathrm{exp}(z_i)}{\sum_j \mathrm{exp}(z_j)}.
\end{equation}

The softmax output is used in audio classification tasks where a test audio is assigned a single label, for example \textit{acoustic scene classification} or \textit{environmental sound classification}. 

In \textbf{multi-class multi-label classification}, multiple classes may be present in an output simultaneously. In this case, softmax activation is replaced by a separate sigmoid activation function for each output, producing probability $\hat{y}_i=P(y=i|\bm{x})$ for each class $i$, and these probabilities do not need to sum to one. This structure is used for \textit{audio tagging} and \textit{sound event detection}, with the difference between them being that the probabilities provided by the outputs are interpreted at different time scale: in audio tagging, a single binary decision is done for each class over the entire duration of the test audio clip, while in sound event detection a binary decision for each class is done in consecutive segments, allowing it to change between 0 and 1 throughout the test audio clip. 

\textbf{Convolutional neural networks} are a specialized kind of networks where not all input and output nodes are connected to each other, but restricted by a kernel. 
In audio processing, the convolution is applied to a time-frequency representation presented as two-dimensional data matrix $X$, using a two-dimensional kernel matrix $K$.
The kernel is usually much smaller than the input feature matrix, for example 3 by 3. The 2D-convolution operation calculates output matrix $S$ as 
\begin{equation}
S(i,j)=(X*K)(i,j)=  \sum_{m} \sum_{n}{X(m,n)K(i-m,j-n)}.
\end{equation}
A typical convolution block consists of four operations. First, several convolutions are performed in parallel. Second, the output of these convolutions go through a nonlinear activation function that acts as a detector and provides a nonlinear transformation. The purpose of this transformation is to allow the network to learn a different feature space, in which the network can represent the solution
Third, the output is further modified by a so-called \textit{pooling function} which replaces the network output at a certain location with a summary statistic of the nearby outputs. The most used pooling is \textit{max pooling}, which is simply the maximum output within a rectangular neighborhood. The role of the pooling operation is to make the representation invariant to small translations of the input. Finally, the representation may be subsampled in time and frequency to reduce its dimensionality. For a complete presentation of CNNs we refer the reader to \cite [Chapter 9] {Goodfellow-et-al-2016}.

Because sound event detection requires estimating temporal activities of sounds, CNNs for SED often use only frequency pooling and do not subsample the representation in time, in order to provide decisions for consecutive segments of the audio clip. For audio classification, a single decision per audio clip is generally required, therefore it is not necessary to preserve the time axis, which allows use of pooling and subsampling in both frequency and time.

\textbf{Recurrent neural networks} (RNN) are specialized for processing sequential data, and therefore highly useful for audio processing. The recurrence means that the output of the network at time $t$ depends also on the hidden state of the network at time $t-1$. The hidden state of the network at time $t$ is defined as: 
\begin{equation}
    \bm{h}^{(t)} = f (\bm{h}^{(t-1)}, \bm{x}^{(t)}; \bm{\theta})
\end{equation}
The output at time $t$ then uses information from the hidden units at time $t$ to make the prediction. This kind of structure allows online operation, where information about past inputs are used in making the prediction at each time step. In offline processing, it is possible to include bidirectional processing that also uses information about future time steps.
The most effective models used in practical applications are \textit{gated} RNNs, which have connection weights that may change at each time step. For example the \textit{long short-term memory} (LSTM) introduces internal recurrence as \textit{gated self-loops}, to produce paths where the gradient can flow for long durations. An LSTM unit has a self-loop controlled by a \textit{forget gate}, controlled by another hidden unit, which can shut off the output of the unit, effectively "forgetting" the old state, while the state is updated with a conditional self-loop weight. Gated recurrent units (GRU) are similar, but have a single gating unit that controls both the forgetting factor and the decision to update the state unit \cite [p.397-401] {Goodfellow-et-al-2016}. 

\textbf{Convolutional recurrent neural networks} (CRNN) combine convolutional and recurrent layers into a single architecture, suitable for tasks in which temporal sequence modeling is advantageous, such as is the case for sound event detection.
The convolutional layers act as feature extractors, aiming to learn discriminative features through the consecutive convolutions and non-linear transformations applied to the time-frequency representation presented at the input of the network, whereas the recurrent layers aim to learn the temporal evolution of the signal.

\end{document}